\documentstyle[12pt]{article}
\jot = 1.5ex\catcode`\@=11\renewcommand{\thefootnote}{\fnsymbol{footnote}}

\def\be{\begin{equation}}
\def\ee{\end{equation}}
\def\l{\label}
\def\F{{\cal F}}

\begin{document}\begin{titlepage}

\rightline{hep-th/9712109}
\rightline{DFPD97/TH/56}

\vspace{1.cm}

\centerline{\large{\bf N=2 SYM RG Scale as Modulus for  WDVV 
Equations}}

\vspace{.5cm}

\vspace{1.5cm}

{\centerline{\sc Gaetano BERTOLDI and Marco MATONE}}

\vspace{0.3cm}

\centerline{\it Department of Physics ``G. Galilei'' and INFN, Padova, 
Italy}
\centerline{bertoldi@padova.infn.it  matone@padova.infn.it}

\vspace{2cm}

\centerline{\bf ABSTRACT}

\vspace{.6cm}
\noindent
We derive a new set of WDVV equations for $N=2$ SYM in which the 
renormalization scale $\Lambda$ is identified with the distinguished 
modulus which naturally arises in topological field theories.

\vspace{0.6cm}
\noindent

\end{titlepage}

\newpage
\setcounter{footnote}{0}
\renewcommand{\thefootnote}{\arabic{footnote}}

A fascinating aspect of Seiberg--Witten theory \cite{SW1} is that it 
relies on deep
geometrical aspects such as uniformization theory \cite{BMT}. 
On the other hand uniformization theory 
turns out to be 
relevant also for topological field theories and non--perturbative
2D quantum gravity 
\cite{UniforPoinca} to which Seiberg--Witten theory is related.
Furthermore, as a consequence of the  Seiberg--Witten results, it 
has been possible
to derive the explicit expression for the $beta$--function 
\cite{Loro}\cite{Noi}\cite{Noi2} which has been 
recently reconsidered 
in 
\cite{Essi1}--\cite{Essi4}. 
In \cite{GIMA}\cite{GBMM}
the analogue of the Zamolodchikov c--theorem
\cite{zam} in the framework of $4D$ N=2 SYM has been considered
(see also \cite{Essi1}\cite{Essi2}\cite{Essi3}).

In the following we will consider the $SU(n)$ gauge group
and denote by $\F$ the prepotential and 
by\footnote{We will denote by $\hat \imath, 
\hat \jmath, \hat k, \ldots$ the
indices running from $1$ to $n-1$ and by $i,j,k,\ldots$ those 
running from $0$ to $n-1$, where $\F_0=\partial_\Lambda\F$.}
$\tau_{\hat\imath\hat\jmath}=\partial^2\F/\partial a^{\hat i}\partial a^{\hat j}$
the effective couplings where
$a^{\hat \imath}= 
\langle \phi^{\hat \imath}\rangle$ are
the vev's of the scalar component.

Let us consider the 
$beta$--function (matrix)\be
\beta_{\hat \imath \hat \jmath}
=\Lambda{\partial \tau_{\hat \imath \hat \jmath}\over \partial \Lambda}
|_{u^2,u^3,\ldots},
\l{betann}\ee
where $u^\alpha\equiv u=\langle {\rm tr\, \phi^\alpha}\rangle$,
$\alpha=2,\ldots,n$.

Very recently, it has been shown in \cite{GBMM} that
$\beta_{\hat \imath \hat \jmath}$
plays the role of the metric whose inverse satisfies the 
WDVV--like equations. In doing this we first considered the WDVV--like 
equations derived in the framework of the $SU(3)$ reduced Picard--Fuchs
equations \cite{KLT}\cite{extension}. Next, we used
the WDVV--like equations derived for $SU(n)$, $n\geq 4$ in 
\cite{MAMIMO}. In \cite{GBMM} it has been observed 
that by Euler theorem 
\be
\beta_{\hat \imath \hat \jmath}=-\Delta^{\hat k}
{\F}_{\hat k \hat \imath \hat \jmath},  
\l{iosqeddddu}\ee
where
\be
\Delta^{\hat k}=\Delta_{u^\gamma}a^{\hat k}= 
\sum_{\gamma=2}^n \gamma u^\gamma{\partial a^{\hat k}
\over \partial u^\gamma}.
\l{scalingopcona}\ee
This strongly
indicates, as should be for a topological
field theory \cite{topologici}, there is actually a natural 
distinguished modulus: 
the renormalization
scale $\Lambda$. This strongly suggests considering the 
extension of the
WDVV--like equation to fully topological WDVV--equations. 
The idea is to
look for WDVV--equations where the derivatives of the
 prepotential
include that with respect to the distinguished
 modulus $\Lambda$.
We will show that there are remarkable relations 
which actually
lead to such a result.

The result in \cite{GBMM} is that the $\beta$--function 
of $N=2$ SYM
with gauge group $SU(n)$,
satisfies the WDVV--like equations
\be
{\F}_{ \hat \imath \hat k \hat l}\beta^{\hat l \hat m} {\F}_{\hat m
\hat n \hat \jmath}=
{\F}_{\hat \jmath \hat k \hat l}\beta^{\hat l \hat m}{\F}_{\hat m
\hat n \hat \imath},
\l{disarmingresult}\ee
where $\beta^{\hat l \hat m}$ is the inverse of the matrix
$\beta_{\hat l \hat m}$ and
$$
\F_{\hat \imath_1\ldots \hat \imath_k}=
{\partial^k \F \over \partial a^{\hat \imath_1}\ldots
 \partial a^{\hat \imath_k}},
$$
$\hat \imath_1,\ldots, 
\hat \imath_k=1,\ldots,n-1$.

Let us write down the WDVV--like equations derived 
in \cite{MAMIMO}
\be
F_{\hat \imath} F_{\hat k}^{-1} F_{\hat \jmath}
= F_{\hat \jmath} F_{\hat k}^{-1} F_{\hat \imath},
\l{WDVV1}\ee
where $(F_{\hat k})_{\hat \imath \hat \jmath}
=\F_{\hat k\hat \imath \hat \jmath}$. We want 
to show that 
\be
F_{ i} F_{k}^{-1} F_{j}
= F_{j} F_{k}^{-1} F_{i},
\l{WDVV2x}\ee
where $(F_{ k})_{ij}=\F_{kij}$ and all the indices run from 
0 to $n-1$ and $a^0 \equiv \Lambda$. 
Observe that this is not a trivial extension
as we also have extended range of summation. Writing
explicitely the summation indices, Eq.(\ref{WDVV2x}) is
\be
({\F_i})_{kl}{(\F_p)}^{lm} ({\F_j})_{mn}=
({\F_j})_{kl}{(\F_p)}^{lm}({\F_i})_{mn},
\l{qdisarmingresult}\ee
where
${(\F_p)}^{lm}$ denotes the inverse 
of the matrix 
$(\F_p)_{lm}=\F_{plm}$.

Observe that
we can define the
 generalized
beta function by
\be
\beta_{ij}=a^0{\partial \F_{ij}\over 
\partial a^0}|_{u^2,u^3,\ldots},
\l{assius}\ee
$i,j=0,\ldots,n-1$. Generalizing the $beta$--function
$\beta^{(a)}$ evaluated in \cite{Noi}, we can introduce
\be
\beta_{ij}^{(a)}=a^0{\partial \F_{ij}\over 
\partial a^0}|_{a^1,a^2,\ldots},
\l{assiuser}\ee

A key point in the derivation of (\ref{qdisarmingresult})
 is the following identity
$$
\F_{ijk}=
$$
$$
\F_{ijk}(1-\delta_{i0})(1-\delta_{j0})
(1-\delta_{k0}) +
$$
$$
-a^{0^{-1}} a^{\hat l} \left[\F_{{\hat l} ij}(1-\delta_{i0})
(1-\delta_{j0})\delta_{k0} + \F_{{\hat l} ik}(1-\delta_{i0})
(1-\delta_{k0})\delta_{j0} + \F_{{\hat l} jk}(1-\delta_{j0})
(1-\delta_{k0})\delta_{i0}\right] +
$$
$$
a^{0^{-2}} a^{\hat l}  a^{\hat m} 
\left[\F_{{\hat l} {\hat m} i}(1-\delta_{i0})\delta_{j0}
\delta_{k0} + 
\F_{{\hat l} {\hat m}j}\delta_{i0}
(1-\delta_{j0})\delta_{k0}) +\F_{{\hat l} {\hat m} k}
\delta_{i0}\delta_{j0}(1-\delta_{k0}) \right)+ 
$$
\be
-a^{0^{-3}}a^{\hat l} a^{\hat m} a^{\hat n} 
\F_{{\hat l} {\hat m} {\hat n}}\delta_{i0}\delta_{j0}\delta_{k0},
\l{cassius}\ee
where the summation on the indices ${\hat l}, {\hat m}, {\hat n} $
goes from $1$ to $n-1$. Observe that on the right hand side of
Eq.(\ref{cassius}) there never appears $\F$ derived with respect
to $a^0$ as these possible terms are killed by the delta 
functions.

The identity (\ref{cassius}) 
is obtained by first observing that by Euler theorem
\be
a^j\partial_{a^j} \F_{i_1,\ldots,i_k} =(2-k)\F_{i_1,\ldots,i_k}. 
\l{cj}\ee
Therefore, for example
\be
\F_i= \F_i(1-\delta_{i0})+a^{0^{-1}}(2\F-a^{\hat k}
\F_{\hat k})\delta_{i0}.
\l{i}\ee
Observe that in this way we get an expression for the 
derivatives of $\F$ with respect
to any $a^k$, including $a^0$, in which only the 
derivatives
of $\F$ with respect to $a^{\hat k}$ appear. 
In this way
 we can express the Eq.(\ref{qdisarmingresult}) 
in terms of $\F$ and its 
derivatives with respect
to $a^{\hat k}$.

In the following we will denote by $\hat F_i$ the matrix obtained from
$F_i$ with the matrix indices running from $1$ to $n-1$. In
other words $(\hat F_i)_{\hat\jmath\hat k}=\F_{i\hat\jmath\hat k}$. 
Similarly,
by $\hat F^{-1}_i$ we will denote the matrix obtained from 
$F^{-1}_i$ by restricting the range of the matrix indices to $1,\ldots,n-1$.
We also define $G_i$ the matrix inverse of $\hat F_i$ (observe that
$G_i\ne \hat F^{-1}_i$).  
Let us derive the expression for the inverse of the
$F_i$ matrix. We set
\be
\rho_i=(F_i)_{00},\qquad (\sigma_i)_{\hat \jmath}=
(F_i)_{0\hat \jmath},
\l{astro1}\ee
\be
\mu_i=(F_i)^{-1}_{00},\qquad (\nu_i)_{\hat \jmath}=
(F_i)^{-1}_{0\hat \jmath}.
\l{astro2}\ee
We have 
$$
\mu_i \rho_i+{}^t\sigma_i \nu_i=1,\qquad \hat F_i
\nu_i+\mu_i\sigma_i=0,
$$
for $i=0,\ldots, n-1$, so that
\be
\mu_i={1\over \rho_i-{}^t\sigma_iG_i\sigma_i},
\l{ahoo}\ee
and
\be
\nu_i= -\mu_iG_i\sigma_i.
\l{ahoo2}\ee
Furthermore, we have
\be
{\hat F_i}^{-1} = \left(G_i\left({\bf 1} - 
\sigma_i \otimes \nu_i \right)\right)_{\hat \jmath \hat k}.
\l{ahoo3}\ee
Therefore, we now have $F_i^{-1}$ in terms of $\rho_i,\sigma_i$ and
$G_i$.

In terms of $\rho_i, \sigma_i$ etc.
Eq.(\ref{qdisarmingresult}) is equivalent to the following 
identities

\vspace{.3cm}

$$
\mu_p \rho_i \rho_j  + \rho_j {}^t\sigma_i \nu_p + 
\rho_i {}^t\nu_p \sigma_j 
+ {}^t\sigma_i {\hat F}_p^{-1} \sigma_j = 
$$
\be 
\mu_p \rho_j \rho_i  + \rho_i {}^t\sigma_j \nu_p + 
\rho_j {}^t\nu_p \sigma_i 
+ {}^t\sigma_j {\hat F}_p^{-1} \sigma_i,
\l{cassius10}\ee

\vspace{.3cm}

$$
\rho_j \mu_p \sigma_i + \rho_j \hat F_i \nu_p + \hat F_i 
{\hat F}_p^{-1} \sigma_j
+ (\sigma_i \otimes \nu_p)\sigma_j = 
$$
\be
\rho_i \mu_p \sigma_j + \rho_i \hat F_j \nu_p + 
\hat F_j {\hat F}_p^{-1} \sigma_i
+ (\sigma_j \otimes \nu_p)\sigma_i,
\l{cassius20}\ee

\vspace{.3cm}

$$
\hat F_i {\hat F}_p^{-1} \hat F_j +(\sigma_i \otimes \nu_p)
\hat F_j + (\mu_p \sigma_i + \hat F_i \nu_p) \otimes
\sigma_j = 
$$
\be
\hat F_j {\hat F}_p^{-1} \hat F_i +
(\sigma_j \otimes \nu_p) \hat F_i + (\mu_p \sigma_j + 
\hat F_j \nu_p) \otimes
\sigma_i.
\l{cassius30}\ee

\vspace{.3cm}

While we immediately see that (\ref{cassius10}) holds, 
the other two equations (\ref{cassius20}) and (\ref{cassius30}) 
are satisfied as a consequence 
of the identity
\be
\hat F_i G_j \sigma_k = \hat F_k G_j\sigma_i, 
\l{cassius40}\ee
where $i,j,k=0,\ldots n-1,$ which follows from 
(\ref{WDVV1})(\ref{cassius}) 
and (\ref{astro1}).
Actually we have
$$
(\hat F_i G_j \sigma_k )_{\hat l}= (\hat F_i)_{\hat l \hat m} 
(G_j)^{\hat m \hat p} 
\left(-{a^0}^{-1}a^{\hat s} \F_{ \hat s k \hat p}(1 -\delta_{k0}) 
+ {a^0}^{-2}a^{\hat s}a^{\hat t} \F_{\hat s\hat t \hat p}
 \delta_{k0}\right).
$$
Then if both $i$ and $k$ are different from 0, the 
identity follows directly from (\ref{WDVV1}), whereas if either
$i$ or $k$ is 
equal to 0, then we simply use the fact that by (\ref{cj})
$$
(\hat F_0)_{\hat \imath \hat k}=-{a^0}^{-1}a^{\hat m}
(\hat F_{\hat m})_{\hat \imath \hat k},
$$
to reduce to previous case.

To see how the identity  works, we give the example
of Eq.(\ref{cassius20}) 
that by (\ref{ahoo2}) and (\ref{ahoo3}) is equivalent to 
$$
\mu_p \left[ \rho_j \sigma_i -\rho_j \hat F_i G_p \sigma_p + \hat F_i 
G_p(\sigma_p \otimes G_p \sigma_p) \sigma_j
- (\sigma_i \otimes G_p \sigma_p)\sigma_j \right. + 
$$
$$
\left. - \rho_i \sigma_j -\rho_i \hat F_j G_p \sigma_p + \hat F_j 
G_p(\sigma_p \otimes G_p \sigma_p) \sigma_i
- (\sigma_j \otimes G_p \sigma_p)\sigma_i \right] + 
$$
\be
\hat F_i G_p\sigma_j - \hat F_j G_p \sigma_i = 0,
\l{maremmascorp}\ee
which is satisfied thanks to Eq.(\ref{cassius40}).

In conclusion, we have derived a new set of WDVV equations for
 $N=2$ SYM, in which
the renormalization scale $\Lambda$ is 
identified with the distinguished modulus which naturally arises
in topological field theories \cite{topologici}.
In particular, repeating the construction in \cite{GBMM}, we
have that Eq.(\ref{qdisarmingresult}) implies 
\be
({\F_i})_{kl}\beta^{lm} ({\F_j})_{mn}=
({\F_j})_{kl}\beta^{lm}({\F_i})_{mn},
\l{qdisarmingresultxxx}\ee
where $\beta^{lm}$ is the inverse of the generalized $beta$ matrix 
$\beta_{lm}$ defined in (\ref{assius}).
Furthermore, it is easy to see that 
\be
({\F_i})_{kl}\beta^{(a)^{lm}} ({\F_j})_{mn}=
({\F_j})_{kl}\beta^{(a)^{lm}}({\F_i})_{mn},
\l{qdisarmingresultxxxyt}\ee
where $\beta^{(a)^{lm}}$ is the inverse of the $beta$ matrix
defined in (\ref{assiuser}).

Finally, we stress that it would be desirable to understand
the structure of the equations for $\F$ deriving from the conditions
of flatness of the WDVV--metrics $\beta_{{lm}}$  and 
$\beta_{{lm}}^{(a)}$. 

\vspace{1cm}

\noindent
{\bf Acknowledgements}. It is a pleasure to thank
G. Bonelli, R. Carroll, J. Isidro, P. Pasti and M. Tonin for discussions.
MM was  supported in part by 
the European Commission TMR programme ERBFMRX--CT96--0045.

\end{document}